\documentclass[usegraphicx,usenatbib]{mn2e} 
\usepackage{epsfig}
\usepackage{ulem}
\usepackage{color}
\usepackage{graphics,graphicx}
\usepackage{times} 
\usepackage{amssymb}
\usepackage{amsmath}
\usepackage{url}
\usepackage{multirow}
\usepackage{ctable}
\usepackage{threeparttable}
\newif\ifAMStwofonts
\AMStwofontstrue
\def\xmm{{\it XMM-Newton~\/}}

\def\rxte{{\it RXTE~\/}}

\def\suzaku{{\it Suzaku}}

\def\epicpn{{\it EPIC}{\rm-pn}}
\def\epicmos1{{\it EPIC}{\rm-MOS1~\/}}
\def\epicmos2{{\it EPIC}{\rm-MOS2 ~\/}}
\def\epicmos{{\it EPIC}{\rm-MOS}}
\def\xis{{\rm XIS}}
\def\pin{{\rm PIN}}

\def\xmm{{\it XMM-Newton}}

\def\rxte{{\it RXTE}}

\def\xspec{\hbox{\sc XSPEC}}

\def\heasoftv{\hbox{\rm HEASOFT\thinspace v6.6.1}}
\def\xselect{\hbox{\rm XSELECT}}
\def\ftool{\hbox{\rm FTOOL}}

\def\s{\hbox{$\rm\thinspace s$}}
\def\ks{\hbox{$\rm\thinspace ks$}}

\def\deg{$^{\circ}$}  

\def\kpc{\hbox{$\rm\thinspace kpc$}}

\def\pcmsq{\hbox{$\rm\thinspace cm^{-2}$}}

\def\ev{\hbox{$\rm\thinspace eV$}}
\def\kev{\hbox{$\rm\thinspace keV$}}

\def\ctsps{\hbox{$\rm\thinspace count~s^{-1}$}}

\def\ergpcmsqps{\hbox{$\rm\thinspace erg~cm^{-2}~s^{-1}$}}

\def\ergps{\hbox{$\rm\thinspace erg~s^{-1}$}}

\def\msun{\hbox{$\rm\thinspace M_{\odot}$}}

\def\rg{${\it r}_{\rm g}$}
\def\rin{${\it r}_{\rm in}$}

\def\refbhb{\rm{\sc REFBHB}}
\def\nh{${\it N}_{\rm H}$}
\def\ka{$K\alpha$}

\def\chisq{{\chi^{2}}}

\def\laor{\rm{\small LAOR~\/}}
\def\phabs{\rm{\small PHABS~\/}}

\def\diskbb{\rm{\small DISKBB~\/}}
\def\compps{\rm{\small compPS~\/}}
\def\comptt{\rm{\small compTT~\/}}

\def\kdblur{\rm{\small KDBLUR}}
\def\kerrconv{\rm{\small KERRCONV}}

\def\xspec{\hbox{\small XSPEC~\/}}

\def\heasoftv{\hbox{\rm{\small HEASOFT}~v6.9\/}}
\def\xselect{\hbox{\rm{\small XSELECT~\/}}}

\def\ftool{\hbox{\rm{\small FTOOL}}}

\def\grppha{\hbox{\rm{\small GRPPHA~\/}}}

\def\mathpha{\hbox{\rm{\small MATHPHA}}}
\def\addascaspec{\hbox{\rm{\small ADDASCASPEC~\/}}}

\def\rmfgen{\hbox{\rm{\small RMFGEN}}}
\def\arfgen{\hbox{\rm{\small ARFGEN}}}

\def\hxddtcor{\hbox{\rm{\small HXDDTCOR}}}

\def\addascaspec{\hbox{\rm{\small ADDASCASPEC}}}

\def\grid25{\hbox{\rm{\small GRID25}}}
\def\aeattcor{\hbox{\rm{\small AEATTCOR}}}
\def\xiscoord{\hbox{\rm{\small XISCOORD}}}
\def\pharbn{\hbox{\rm{\small PHARBN}}}
\def\epatplot{\hbox{\rm{\small EPATPLOT}}}

\def\gx{\hbox{\rm GX 339-4}}
\def\jb{\hbox{\rm J1655-40}}
\def\j17{\hbox{\rm J1753.5-0127}}
\def\j118{\hbox{\rm XTE J1118+480}}

\def\j{\hbox{\rm XTE J1752-223}}
\def\cyg{\hbox{\rm Cyg~X-1}}

\begin{document}

\title[Spin from reflection features: XTE~J1752-223] {Multi-state observations of the Galactic Black Hole \j: Evidence for an intermediate black hole spin.}  \author[R. C. Reis et al.]
{\parbox[]{6.in} {R.~C.~Reis $^{1}$\thanks{E-mail:
      rcr36@ast.cam.ac.uk}, J.~M.~Miller$^{2}$, A.~C.~Fabian$^{1}$, E.~M.~Cackett$^2$, D.~Maitra$^2$, C.~S.~Reynolds$^{3,4}$, M.~Rupen$^5$, D.T.H.~Steeghs$^6$, and R.~Wijnands$^7$  \\}\\
  \footnotesize
  $^{1}$Institute of Astronomy, Madingley Road, Cambridge, CB3 0HA\\ 
   $^{2}$Department of Astronomy, University of Michigan, 500 Church Street, Ann Arbor, MI 48109, USA\\
    $^{3}$Department of Astronomy, University of Maryland, College Park, MD, 20742, USA\\
     $^4$Joint Space Science Institute (JSI), University of Maryland, College Park, MD20742, USA\\
     $^5${\it NRAO}, Array Operations Center, 1003 Lopezville Road, Socorro, NM 87801, USA\\
     $^6$Department of Physics, University of Warwick, Coventry CV4 7AL\\
     $^7$Astronomical Institute ``Anton Pannekoek'', University of Amsterdam, Postbus 94249, 1090 GE Amsterdam, the Netherlands
     }

\maketitle

\begin{abstract} The Galactic Black hole candidate \j\ was observed during the decay of its 2009 outburst with the \suzaku\ and \xmm\ observatories. The observed spectra are consistent with the source being in the ''intermediate`` and ''low-hard state`` respectively.  The presence of a strong, relativistic iron emission line is clearly detected in both observations and the line profiles are found to be remarkably consistent and robust to a variety of continuum models. This strongly points to the compact object in \j\ being a stellar-mass black hole accretor and not a neutron star. Physically-motivated and self-consistent reflection models for the Fe-\ka\ emission-line profile and disk reflection spectrum rule out either a non-rotating, Schwarzchild black hole or a maximally rotating, Kerr black hole at greater than 3$\sigma$ level of confidence. Using a fully relativistic line function in which the black hole spin parameter is a variable, we have formally constrained the spin parameter to be $0.52\pm0.11 (1\sigma)$. Furthermore, we show that the source in the low--hard state still requires an optically--thick disk component having a luminosity which is consistent with the $L\propto T^4$ relation expected for a thin disk extending down to the inner--most stable circular orbit. Our result is in contrast to the prevailing paradigm that the disk is truncated in the low-hard state. 

\end{abstract}


\section{Introduction}
Stellar-mass Galactic black holes in X-ray binaries represent nearby
laboratories in which the inner regions of the accretion flow can be studied in detail, thus providing information on both the geometry
of the accretion disc and on intrinsic physical parameters
such as black hole mass and spin. Whereas the supermassive black holes that power active galactic nuclei (AGN) are variable by a factor of several, the mass accretion rate onto a stellar-mass black hole can vary by $10^{8}$ between quiescence and outburst peak.  This affords unique opportunities to study how accretion flows evolve, including how relativistic jets are produced and quenched (\citealt{fenderetal04}).  At relatively high mass accretion rates, where a standard \citet{shakurasunyaev73} accretion disk remains at the innermost stable circular orbit (ISCO), the X-ray spectrum of these sources permit constraints on black hole spin parameters (e.g. \citealt{miller02j1650, mcclintock06, shafee06, reisgx, miller09spin}). 

Presently, dynamical constraints demand a black hole primary in 20
sources in the Milky Way and Large Magellanic Cloud (\citealt{Remillard06}).  In approximately twice this number of sources,
observing constraints have prevented dynamical constraints on the mass
of the primary, but X-ray spectra, timing  and
multi-wavelength properties clearly indicate that the binary harbours a
black hole.  These sources are often called "black hole candidates"
(\citealt{Remillard06}).  New candidate black hole transients
that are found to lie in favourable fields and to not suffer excessive
extinction -- candidates that will permit dynamical constraints via radial
velocity curves -- are especially important sources.

XTE J1752$-$223 was discovered during periodic \rxte\ scans of the
Galactic bulge region on 23 October 2009 (\citealt{1752}).  Observations with the {\it Swift}/BAT and {\it MAXI}/GSC soon confirmed the new source (\citealt{1752b, nakahira09}).  Additional prompt
observations failed to detect pulsations; the absence of pulsations
and the nature of the spectrum led to an early identification of XTE
J1752$-$223 as a new black hole candidate (\citealt{1752b}).  Many black hole X-ray binaries begin their rise to
outburst peak in the ``low/hard'' state, in which jet emission is
typical (\citealt{Remillard06}); prompt radio
observations of XTE J1752$-$223 with {\it ATCA} detected a
flat-spectrum radio sources, consistent with emission from a compact
jet (\citealt{brocksopp09}).  Also typical of black
holes, XTE J1725$-$223 was detected in hard X-rays with the {\it
Fermi}/GBM (\citealt{Wilson-Hodge09}).  A bright optical counterpart has been reported (\citealt{torres2009a}), together with the presence of a broad H$\alpha$ emission line (\citealt{torres2009b}). The column
density along the line of sight to XTE J1752$-$223 is modest
($4.5\times 10^{21}~{\rm cm}^{-2}$, \citealt{Dickey90}), and a
dynamical mass constraint may be possible.

Understanding the role of black hole spin in shaping accretion flows
onto -- and jets from -- black holes is an important and timely aim.
The dimensionless spin parameter of a black hole is given by $a =
cJ/GM^{2}$, where $J$ is angular momentum, $G$ is Newton's constant,
and $M$ is the mass of the black hole.  In the case of stellar-mass
black holes, the spin parameter cannot be changed significantly
through the limited mass that can be accreted from the companion star
(\citealt{thorne74, gammie04}).  Thus, the spin of
black holes in X-ray binaries is set by the creation event, likely a
gamma-ray burst (GRB) or supernova (SNe).

The mass of the black hole in XTE J1752$-$223 and the distance to this
source are not presently known, but modelling of relativistic Fe K disk
lines -- as well as all other associated reflection features -- in the spectra of X-ray binaries, can constrain spin parameters
without these quantities (e.g. \citealt{miller02j1650, miller04gx, miller09spin,martocchia02, reisgx, reis09spin}). The reflection component arise as hard emission  from the corona irradiates the cooler disc and results in ``reflection signatures'' consisting of fluorescent and
  recombination emission lines as well as absorption features (\citealt{reflionx}). The
  most prominent of these signatures being the broad, skewed Fe-\ka\
  line. Constraint on the spin parameter arises because the
Doppler shifts and gravitational red shifts that shape such lines
depend only on the relative depth of the disk within the potential
well, not on an absolute quantity.  When the mass and distance to a
source are known, modelling of the accretion disk continuum can give a
completely independent spin constraint (e.g. \citealt{mcclintock06, gou09}).

In this paper, we draw on observations of XTE J1752$-$223 with {\it
XMM-Newton} and {\it Suzaku} to constrain the nature of the innermost
accretion flow in this source, and to constrain the spin parameter of
the black hole.  The observations were made at very different points
during the current outburst of XTE J1752$-$223: {\it Suzaku} observed
the source in an "intermediate" state, while {\it XMM-Newton} observed
the source in a "low/hard" state.  A relativistic iron disk line is
detected in both observations; the profile is remarkably consistent
between the two.

\section{Data reduction}
\label{observation}

\begin{figure}
\centering
{
 \rotatebox{0}{
{\includegraphics[width=8.5cm]{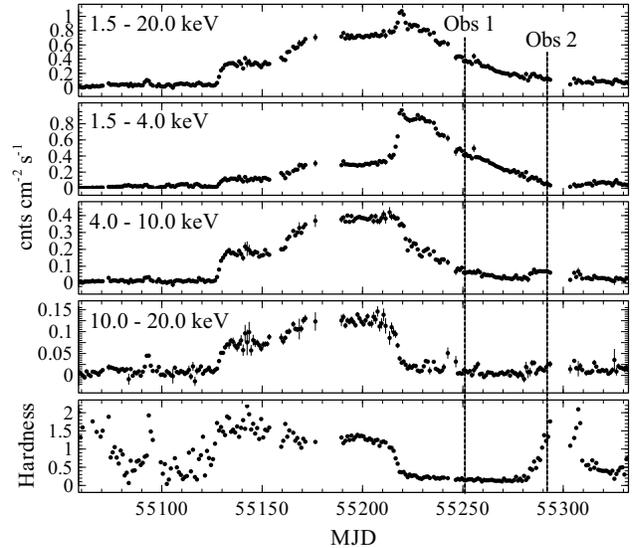}  
}}}

\caption{MAXI lightcurve for \j\ in four energy bins and hardness ratio (Bottom panel) defined as the ratio between the 4--10 and 1.5--4\kev\ energy range. The dotted vertical lines shows the time for the \suzaku\ (Obs.~1) and \xmm\ (Obs.~2) observations presented in this paper. It is clear that the source was observed in two different spectral states. }
\label{fig_lightcurve}
\end{figure}

XTE ~J1752-223 was observed during the decay of its 2009 outburst with \suzaku\ (\citealt{SUZAKU}) on February 24 (hereafter Obs.~1), and subsequently with \xmm\  (\citealt{XMM}) on April 06 (hereafter Obs.~2) for a total exposure of approximately 42\ks\ each time. The \suzaku\ observation caught the source in the intermediate state during its decay from maximum flux as shown in Fig.~\ref{fig_lightcurve}. At the time of the  \xmm\ observation, \j\  was back in the low-hard state.

The three operating detectors constituting the X-ray Imaging Spectrometer (XIS; \citealt{SUZ_XIS}) on-board of \suzaku\ were operated in the ``burst'' mode with the front illuminated (FI; XIS0 and 3) and back illuminated (BI; XIS1) detectors in the 2x2 and 3x3 editing modes respectively. The observation resulted in a total (co-added) good-exposure of approximately 12.2\ks\ and 516\s\ for the FI and BI instruments respectively. Due to the short exposure time of the BI camera, throughout this paper we will only discuss the results obtained with the FI instrument. Using the latest \heasoftv\ software package we processed the unfiltered event files for each of the two remaining CCDs following the \suzaku\ Data Reduction Guide\footnote{http://heasarc.gsfc.nasa.gov/docs/suzaku/analysis/}. New attitude files were created using the \aeattcor\ script\footnote{http://space.mit.edu/cxc/software/suzaku/aeatt.html} (\citealt{aeattcor}) in  order to correct for shift in the mean position of the source caused by the wobbling of the optical axis.  The \ftool\ \xiscoord\ was used to create new event files which were then further corrected by re-running the \suzaku\ pipeline with the latest calibration, as well as the associated screening criteria files. The good time intervals provided by the \xis\ team were employed in all cases to exclude  any possible telemetry saturations.  \xselect was used to extract spectral products from these event files. Source events were extracted from a square-annulus region with inner and outer width of 30 and 240\arcsec\ respectively, and background spectra from another region of the same size, devoid of any obvious contaminating emission, elsewhere on the same chip. The script ``xisresp''{\footnote {http://suzaku.gsfc.nasa.gov/docs/suzaku/analysis/xisresp}} with the ``medium'' input was used to obtain individual ancillary response files (arfs) and redistribution matrix files (rmfs). ``xisresp'' calls the tools ``xisrmfgen'' and ``xissimarfgen''. 
Finally, we combined the spectra and response files from the two front-illuminated instruments (XIS0 and XIS3) using the \ftool\ \addascaspec\ to increase signal-to-noise.  The \ftool\ \grppha\ was used to give at least 100 counts per spectral bin. The Hard X-ray Detector (HXD; \citealt{SUZ_HXD}) was operated in the normal
mode. The appropriate response and tuned  non-X-ray background (NXB) files for HXD-nominal
pointing were downloaded{\footnote{http://www.astro.isas.ac.jp/suzaku/analysis/hxd/}}
and the data were reprocessed in accordance with the \suzaku\ Data
Reduction Guide. Common good time intervals were obtained with MGTIME which combines the good times of the event and background files, and \xselect\ was used to extract spectral products. Dead time corrections were
applied with \hxddtcor, and the exposures of the NXB spectra
were increased by a factor of ten, as instructed by the data reduction
guide. The contribution from the Cosmic X-ray Background (CXB)  was simulated using the form of \citet{Boldt87}, with the appropriate normalisation for
the HXD nominal pointing, resulting in a CXB rate of $0.021\ctsps$.
The NXB and CXB spectra were then combined using \mathpha\
to give a total background spectrum, to which a 2~per~cent systematic uncertainty was added.  The source spectrum was finally grouped to at least 500 counts per spectral bin.

The \epicpn\ camera (\citealt{XMM_PN}) on-board \xmm\ was
operated in ``timing'' mode with a ``medium'' optical blocking
filter. The \epicmos1\ and \epicmos2\ cameras (\citealt{XMM_MOS}) were operated in the ``imaging'' mode. Starting with the unscreened level 1 data files, we generated concatenated and calibrated event lists for the different instruments using the latest \xmm\ {\it Science Analysis System \thinspace v 10.0.0 (SAS)}. \epicpn\ events were extracted from a stripe in RAWX (30-48) {\it vs} RAWY (2.5-199.5) space. Bad pixels and events too close to chip edges were ignored by requiring ``FLAG = 0'' and ``PATTERN $\leq$ 4''. The energy channels were initially binned by a factor of five to create a spectrum. The \epicmos\ data suffered heavily from pile up and will therefore not be included in the forthcoming analysis. We used the SAS task \epatplot\ to assess the level of pile-up in the \epicpn\ spectrum and found it to be insignificant. The \epicpn\ count rate is approximately 150~\ctsps\ which is well below the nominal pile-up limit for the instrument in timing mode ($\approx800~$\ctsps). Response files were created in the standard way using the tools \rmfgen\ and \arfgen. The total good exposure time selected was 1.8\ks. Because of the high source flux in the \epicpn\ spectrum we did not subtract any background. Finally we rebinned the spectrum with the tool \pharbn {\footnote{http://virgo.bitp.kiev.ua/docs/xmm$_{-}$sas/Pawel/reduction/pharbn}}, to have 3 energy channels per resolution element, and at least 20 counts per channel. Throughout this paper XIS-FI and \epicpn\ spectra are fitted in the 1.3--10.0\kev\ energy range respectively. This lower limit was chosen as it was found that the data below this energy showed strong positive residuals above any reasonable continuum. Similar residuals were reported by \citet{hiemstra1652} based on an \epicpn-timing observation of XTE~J1652-453 where the authors associated it with possible calibration issues related with the redistribution matrix of the timing-mode data.   The \pin\ spectrum is restricted to the 20.0--45.0\kev\ energy range and fit simultaneously with the \xis\ data by adding a normalisation factor of 1.18 with respect to that of the FI spectrum. All errors reported in this work are 90~per~cent confidence errors obtained by allowing all parameters to vary, unless otherwise noted.

\section{Data Analyses and Results}

\subsection{Phenomenological models}
\label{simple}

The X-ray spectrum of stellar mass black hole binaries can usually be phenomenologically characterised by a continuum consisting of an absorbed powerlaw together with a thermal disk-blackbody component. In addition to this continuum, often  there is also a broad emission line at $\sim 6.4$\kev\ (see e.g. \citealt{miller07review}). Anticipating a similar combination for the continuum of \j, we start by fitting the energy range  1.3--4.5 and 8.0--45.0\kev\ with a powerlaw modified by interstellar absorption (\phabs{\footnote{ {Using the standard BCMC cross-sections (\citealt{balucinska}) and ANGR abundances (\citealt{abundances}).}}} model in \xspec) together with the disk blackbody model \diskbb\ (\citealt{diskbb, makishima86}). Initially we constrain the neutral-hydrogen column density (\nh) to be the same between the two observations. The model parameters are thus the (global) \nh, as well as disk temperatures ($T_{disk}$), powerlaw indices ($\Gamma$) and normalizations ($N$)  for each observation. This combination resulted in a good description of the continuum, with the bulk of the residuals coming from the energy range between 1.7--2.5\kev\ possibly due to the Au~M-shell edges and Si features in the detectors. Hereafter this energy range will be ignored. The resulting fit gives  $\chi^2/\nu= 2121.5/1282$ with a disk temperature of $0.570\pm0.002$\kev\ and $0.287\pm0.004$\kev\ for Obs.~1 and 2 respectively. The total, luminosity in the 0.5--10\kev\ range, ($L_T$) is $L_{T1}\simeq 2.5(d/10\kpc)\times 10^{38}\ergps$ and $L_{T2}\simeq 3.0(d/10\kpc)\times 10^{37}\ergps$ for Obs.~1 and 2 respectively. Assuming a fiducial black hole of 10\msun\ at a distance of 10\kpc\ places our observations at $L/L_{Edd}\simeq 0.2$ and $0.02$ respectively. The latter value for the luminosity state of \j\ during the \xmm\ observation is over an order of magnitude higher than the range explored by \citet{tomsick09gx} for \gx\ (0.14\%~$L_{Edd}$), where the authors presented the first direct evidence for the truncation of the accretion disk in the low-hard state of accreting black holes.

The unabsorbed disk flux decreases from $(2.57\pm0.02)\times10^{-8}$\ergpcmsqps\ in Obs.~1 to $(1.71\pm0.07)\times10^{-9}$\ergpcmsqps\ in Obs.~2 so that $(T_1/T_2)^4 = 17.7\pm1.0$ and $F_1/F_2 = 15.0\pm1.0$ which is remarkably consistent  with the $L\propto T^4$ relation expected for a standard thin accretion disk extending down to the ISCO (see e.g. \citealt*{fkrbook}) in both spectral states.

\begin{figure}
\centering
{\includegraphics[angle=270, width=7.0cm]{figure_ratio_pl+diskbb.ps} }
{\includegraphics[angle=270, width=7.0cm]{figure_ratio_comptt+diskbb.ps} } 
{\includegraphics[angle=270, width=7.0cm]{figure_ratio_compps+diskbb.ps} }
{\includegraphics[angle=270, width=7.0cm]{figure_ratio_compps2+diskbb.ps} }
\caption{Data/model ratio to a variety of absorbed Compton components together with thermal disk emission. \suzaku\ spectra (\xis\ and \pin) for the Feb-2010 observation are shown in black. The \xmm\ (\epicpn) April-2010 spectrum is shown in red. The \xis\ and \epicpn\ data were fitted in the 1.3--4.5 and 8.0--10.0\kev\ energy range. The \pin\ data was fitted between 20.0--45.0\kev. The bottom panel shows the double Compton model applied to the low-hard state data only. The residuals show that a broad iron-emission line extending to approximately 4\kev\ is present in both observations independent of the model used for the hard emission. The data have been rebinned for plotting purposes. }
\label{fig_ratio2pl}
\end{figure}

\begin{figure}
\centering
{\includegraphics[width=8.5cm, height=5.5cm]{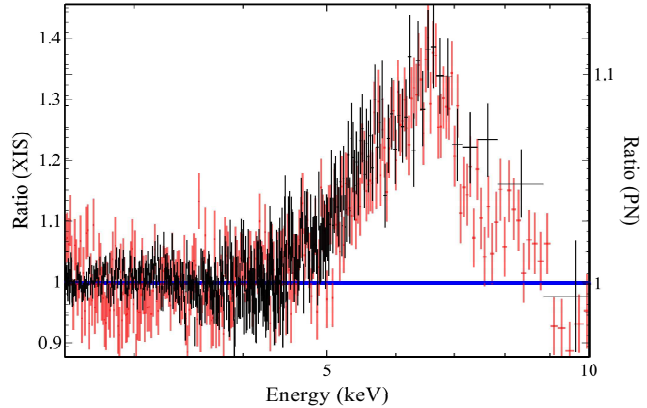}  }
\caption{Close up of the line profile for the \suzaku\ (black) and \xmm\ (red) observations. It is clear that the line profile has varied very little between the observations. Note that the \xmm\ observation has its own ratio scale shown on the right.}
\label{fig_ratio_norm}
\end{figure}

\begin{figure}
\centering
{
 \rotatebox{270}{ 
 \resizebox{!}{8cm} 
{\includegraphics[]{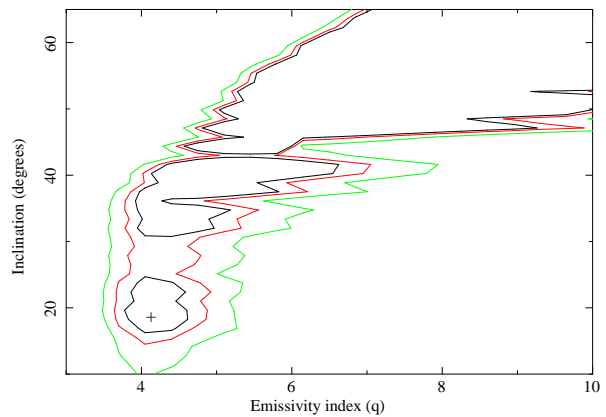}  
}}}
\caption{Inclination versus emissivity index contour plots for \j\ (Model~1). The 68, 90 and 95 per~cent confidence range for two parameters of interest are shown in black, red and green respectively. The cross marks the position of the global minimum.       }
\label{fig_contour}
\end{figure}

The total neutral hydrogen column density in the line of sight to low-mass X-ray binaries is not expected to vary at the time scales considered here (\citealt{nhpaper}). However due to the uncertainties in the cross calibration between the instruments we hereafter allow this parameter to  differ between the \suzaku\ and \xmm\ observations in order to obtain a better description of the continuum. Fig.~\ref{fig_ratio2pl} shows the data/model ratio for the two observations of \j\ fitted with a variety of plausible continuum models over the full energy range allowed for each spectrum excepting the iron-K band between 4.5--8.0\kev.  The models used for the continua are listed in Table~\ref{table}. Model~2 replaces the powerlaw component in Model~1 with the thermal Comptonization code, \comptt, of \citet{comptt} where the input seed photon temperature ($kT_{0}$) was tied to the thermal disk temperature modelled with \diskbb.  The temperature ($KT_e$) and optical depth ($\tau$) of the plasma are further free parameters of this model. As a further test to the robustness of the line profile we replaced in Model~3 the \comptt\ component with \compps\ (\citealt{compps}) which is better suited for high electron temperatures. It is clear from Fig.~\ref{fig_ratio2pl} and Table~\ref{table} that the extent and overall shape of the line does not strongly depend on the chosen continuum.

\begin{table*}
\begin{center}
\caption{Joint fits with simple phenomenological models}
\label{table}
\begin{tabular}{lcccccccccc}                
  \hline
  \hline 
& \multicolumn{2}{c}{Model~1} & \multicolumn{2}{c}{Model~1b}& \multicolumn{2}{c}{Model~2} & \multicolumn{2}{c}{Model~3} \\

& \multicolumn{2}{c}{diskbb+powerlaw} & \multicolumn{2}{c}{diskbb+powerlaw}& \multicolumn{2}{c}{diskbb+compTT} & \multicolumn{2}{c}{diskbb+compPS}\\
Parameters &  Obs.~1   & Obs.~2 & Obs.~1&Obs.~2&Obs.~1 & Obs.~2 & Obs.~1&Obs.~2\\
\nh\ (~$\times10^{22}$\pcmsq) & $0.56\pm0.01$ &$0.14\pm0.02$ & $0.563\pm0.005$ &$0.134\pm0.005$& $0.551^{+0.009}_{-0.004}$ &  $0.15(f)$& $0.550^{+0.008}_{-0.004}$ &$0.07\pm0.01$ \\
$T_{disk}$(\kev)  & $0.561^{+0.002}_{-0.001}$ & $0.42^{+0.02}_{-0.01}$& $0.561^{+0.005}_{-0.001}$ & $0.429^{+0.010}_{-0.004}$ &$0.560^{+0.001}_{-0.002}$ & $0.336^{+0.004}_{-0.005} $ & $0.563\pm0.002$ &$0.434\pm0.001$ \\
$\Gamma$        & $2.57^{+0.06}_{-0.05}$ & $1.82\pm0.01$ & $2.52^{+0.03}_{-0.05}$ & $1.82\pm0.01$&... &...&...&...\\
$kT_{0}$(\kev)  & ... &... & ... &...&  $0.560^{+0.001}_{-0.002}$ & $0.336^{+0.004}_{-0.005}$ &$0.563\pm0.002$ &$0.434\pm0.001$  \\
$kT_{e}$(\kev)  & ... & ...&...&...&$96^{+160}_{-60}$ &$32^{+98}_{-14}$ &$101^{+39}_{-36}$ &$27^{+98}_{-14}$  \\
$\tau$          & ... & ...& ... & ...& $0.1^{+0.8}_{-0.1}$ &$1.2^{+0.7}_{-0.5}$ & $0.3^{+0.7}_{-0.2}$ &$3.0_{-0.8}$  \\
$E_{Laor}$ (\kev) & $6.91^{+0.06}_{-0.21}$ & $6.97_{-0.03}$& $6.97_{-0.23}$ & $6.97_{-0.01}$& $6.84^{+0.17}_{-0.20}$ & $6.97_{-0.02}$& $6.93^{+0.04}_{-0.22}$ & $6.97_{-0.04}$  \\
EW (\ev) & $270^{+70}_{-50}$& $180\pm20$& $280^{+50}_{-40}$& $190\pm20$&  $250\pm50$  &$150\pm10$ & $290\pm60$  & $170^{+30}_{-10}$\\
$q$  & $4.2\pm0.5 $ & $3.3^{+0.3}_{-0.1}$& $4.6\pm0.4 $ & $3.9\pm0.3 $& $4.2^{+0.9}_{-0.5}$ & $3.2\pm0.1$ & $4.3^{+0.5}_{-0.4} $ & $3.2\pm0.1 $\\
$i$ (degrees)  & $19^{+7}_{-3}$& =Obs.~1 & $28\pm2$& =Obs.~1 & $16^{+3}_{-6}$ & =Obs.~1  & $18^{+5}_{-3}$& =Obs.~1 \\
\rin\ (\rg) &$4.1^{+0.3}_{-0.6} $ &  =Obs.~1 &$2.8^{+0.3}_{-0.1} $ &$3.9\pm0.5$ &$4.9\pm0.5$ &=Obs.~1 &$4.0\pm0.4 $ &  =Obs.~1 \\
$\chi^{2}/\nu$& \multicolumn{2}{c}{2645.4/2247}& \multicolumn{2}{c}{2634.9/2246}& \multicolumn{2}{c}{2672.0/2246}& \multicolumn{2}{c}{2626.7/2245}\\
  \hline 
\end{tabular}
\end{center} 
\small Notes: Results of joint \suzaku\ (Obs.~1) and \xmm\ (Obs.~2) fits with simple continuum models. In all cases the inclination and inner disk radii were tied between the observations and the disk component was modelled with the standard multicolour disk model \diskbb\ (\citealt{diskbb}). The hard continuum in Model~1 is assumed to consist of a simple powerlaw. Models~2 and 3 replaces the powerlaw with the Comptonization codes of \citet{comptt} and \citet{compps} respectively. In both cases the input photon temperature ($kT_{0}$) were tied to the thermal disk temperature. For Model~2 a disk geometry was assumed and the column density in Obs.~2 was frozen at the value found in Model~1 so as to prevent it from going to zero. In Model~3 we assume that the Compton cloud has a spherical geometry (parameter\=4) and deactivated the reflection option. All errors are 90~per~cent confidence.
\end{table*}

Allowing the neutral hydrogen column--density to differ between the observations resulted in a higher disk temperature and lower \nh\ for the second observation as compared to the results presented above. Although this fit yields an improvement in $\chisq$ it is unlikely that such a large variation in \nh\ is real and the accompanying high temperature is merely a response to the low \nh. We remind the reader that the purpose of these phenomenological models is simply to highlight the robustness of the line profile to a variety of  -- very different -- continua. From Table~1 we see that the strength of the line -- characterised by its equivalent width -- for Obs.~1~(Obs.~2) does not depend  on the curvature of the continuum, with a value of approximately 270(180)\ev\ for the least curved model (Model~1) and 290(170)\ev\ for the highly curved Comptonisation model \compps (Model~3).

We again see from Fig.~\ref{fig_ratio_norm} that the shape of the line profile does not vary significantly during the two observations even though we are looking at two distinct states of \j\ (see Fig.~\ref{fig_lightcurve}). Such a line profile is usually attributed to  gravitational effects close to the central regions of a black hole (\citealt{Fabian89, laor}). Alternative explanations such as Comptonization, velocity shifts and/or scattering effects have been shown to be extremely unlikely source of line broadening (e.g. \citealt{fabianetal95,reynoldsandwilms2000, milleretal2004Achandra}, see also \citealt{hiemstra1652}). We therefore proceed by assuming a relativistic origin for the broad emission line and as such begin by modelling the residuals seen in Fig.~\ref{fig_ratio2pl} with the the \laor\ model (\citealt{laor}). This model describes a broad line emerging from an accretion disk with emissivity profile described by a power-law of the form $\epsilon_{(r)} = r^{\it -q}$ and an inner radius \rin\ in units of \rg$=GM/c^2$. The outer disc radius was fixed at the maximum allowed value of 400\rg. Only the inner radius and disc inclination, {\it i} were tied between the observations.

\begin{figure}
{
 \rotatebox{270}{ 
 \resizebox{!}{8.cm} 
{\includegraphics[]{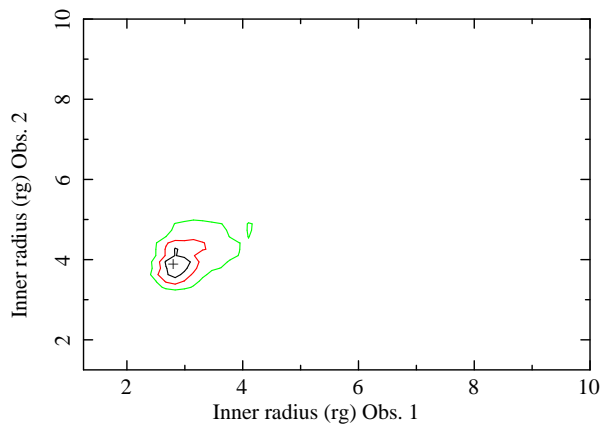}}}}
\caption{Contour plot showing the inner-edge radii for both the intermediate (Obs.~1) and low-hard state (Obs.~2) observations of \j. This provides explicit evidence that the radial extent of the accretion disk does not change between these states. As before the 68, 90 and 95 per~cent confidence range for two parameters of interest are shown in black, red and green respectively. The cross marks the position of the global minima.         }
\label{fig_contour_rin}

\end{figure}

When we fit the line profile with a model such as \laor, together with a separate disk component (e.g. Model~1) we find a degeneracy between the emissivity index ($q$) and the inclination of the system. The inclination as such is determined from the blue wing of the line profile and in \j\ this region is coincidental with the downturn of the disk emission. Figure~\ref{fig_contour} shows the emissivity index\footnote{We are only showing the emissivity index for Obs.~1 versus inclination however, a similar degeneracy is also found for $q$ in Obs.~2.} versus inclination parameter space for Model~1. It is apparent that there are two set of solutions: ({\rm i}) disk inclination $i\lesssim50$\deg\ with $3\lesssim q \lesssim 5$, and ({\rm ii}) $i>50$\deg\ with $q>5$. We will proceed  with the  rest of this analysis by confining the inclination to be less than 40 degrees and $q<5.5$ thus being consistent with values for emissivities found in the literature  (see, e.g., \citealt{miller07review}) and adhering to the value of 3 expected for a standard accretion disk (\citealt{reynoldsnowak03}). It should also be said that we do not expect the inclination of this system to be high based on both the lack of dips and absorption - as expected for edge-on systems - as well as the lack of high frequency QPOs. Furthermore, we will show in the following section that this degeneracy is indeed broken when we fit the data with a self-consistent reflection model that also incorporates the thermal emission. Table~\ref{table} details the various parameters found for the three continuum models together with the \laor\ line profile. In all cases we find a strong relativistic line with an inner radius of approximately 4\rg. In order further to address the question of whether the edge of the inner disk recedes between the two states we allowed in Model~1b for \rin\ to vary between the observations. Figure~\ref{fig_contour_rin} shows the contour map for these two values. It is clear that in {\it both} states the inner-edge of the accretion disk in \j\ extends down to approximately 4\rg, with such a value being consistent amongst the datasets at $2\sigma$ (green contour in Fig.~\ref{fig_contour_rin}).

\begin{table*}
\begin{center}
\caption{ Joint fits with self-consistent reflection models }
\label{table2}
\begin{tabular}{lcccccccc}                
  \hline
  \hline 
& \multicolumn{2}{c}{Model~4} & \multicolumn{2}{c}{Model~5}  \\

& \multicolumn{2}{c}{kdblur*refbhb + powerlaw} & \multicolumn{2}{c}{kerrconv*refbhb + powerlaw} \\
Parameters &  Obs.~1   & Obs.~2 & Obs.~1 & Obs.~2 \\
\nh\ (~$\times10^{22}$\pcmsq) & $0.36\pm0.01$ &$0.34\pm0.01$& $0.37^{+0.01}_{-0.02}$ &$0.36^{+0.05}_{-0.04}$ \\
$\Gamma$ &$2.53^{+0.08}_{-0.09}$ & $1.83\pm0.02$&$2.54^{+0.02}_{-0.11}$ & $1.83\pm0.02$ \\
$T_{disk}$(\kev)  & $0.47^{+0.02}_{-0.01}$ & $0.30^{+0.01}$ & $0.47\pm0.01$ & $0.305\pm0.005$ \\
$F_{Illum}/F_{BB}$ &$0.3^{+0.2}_{-0.1}$& $4.0^{+0.1}_{-0.2}$&$0.28^{+0.05}_{-0.08}$& $3.94^{+0.13}_{-0.15}$\\
$H_{den}$ ($\times10^{20}$) &$0.22^{+0.06}_{-0.07}$ & $25.8^{+3.7}_{-3.0}$&$0.22^{+0.06}_{-0.02}$ & $26^{+3}_{-2}$ \\
$N_{Refbhb}$ ($\times10^{-2}$) &$44^{+16}_{-19}$  &$4.4^{+0.8}_{-0.7} $ &$45^{+5}_{-1}$  &$4.5^{+0.2}_{-0.4} $\\
$N_{hard}$ & $0.12^{+0.12}_{-0.11}$&$0.24\pm0.01$& $0.13^{+0.04}_{-0.13}$&$0.24\pm0.01$ \\
$q$ &$3.6^{+1.4}_{-0.5}$ & $2.4^{+0.2}_{-0.1}$ &$4.0^{+1.1}_{-0.4}$ & $2.5\pm0.1$\\
$\theta$ (degrees) & $25^{+7}_{-8}$ & =Obs.~1& $28^{+4}_{-8}$ & =Obs.~1 \\
\rin\ (\rg) &$3.7^{+0.6}_{-0.7}$ &=Obs.~1&...&... \\
Spin ($a$) &...& ... &$0.52^{+0.13}_{-0.16}$ &=Obs.~1 \\

$\chi^{2}/\nu$& \multicolumn{2}{c}{2579.3/2247} & \multicolumn{2}{c}{2580.2/2247}\\
\hline
\hline
\end{tabular}
\end{center} 
\small Notes: Results of joint \suzaku\ (Obs.~1) and \xmm\ (Obs.~2) fits with the self-consistent reflection model \refbhb. In all cases the inclination, inner disk radii (or spin) were tied between the observations. Model~4 uses the kernel from the \laor\ line profile to account for the gravitational effects close to the black hole.  Model~5 replaces \kdblur\ with the fully relativistic code \kerrconv\ were the spin is a parameter of the model. In both cases the hard emission illuminating the disk is assumed to be a powerlaw with index $\Gamma$. All errors are 90~per~cent confidence for one parameter.
\end{table*}

\suzaku\ observations of the stellar mass black holes \jb\ and \cyg\ in the low-hard state have suggested that the the broad-band continuum in this state has a more concave shape than that predicted by a single \compps\ component together with reflection (\citealt{takahashi165508,makishimacygx108}). To account for this effect the authors proposed a double-\compps model in line with previous double-Comptonisation work (\citealt{gierlinski97,ibragimov05}). Such a model is suggestive of a system where the accretion disk is truncated and the Comptonisation region is inhomogeneous (for a detailed description of this model and interpretation see \citealt{takahashi165508}). In order to test whether such curvature in the continuum could result in a narrower line profile in the low-hard state observation of \j\ we include an extra \compps component to Model~3. Similarly to the work of \citet{makishimacygx108} we constrain both components to have the same seed-photon and electron temperature, and the same reflection parameters, but allow them to differ in
normalization and optical depth. The electron temperature could not be constrained and was thus frozen at 100\kev\ similarly to the value found for \cyg. The presence of a strong and broad emission line (again modelled with the \laor\ line profile) at $6.7\pm0.1$\kev\ remained nonetheless (see bottom panel of Fig.~\ref{fig_ratio2pl}), with an equivalent width of $130^{+20}_{-10}$\ev\ slightly lower than the values found for all other combinations for the continuum (see Table~\ref{table}) as expected for such a curved continuum . However, this fit to the low-hard state dataset ($\chisq/\nu=1162.8/1566$) still requires an accretion disk extending to within $3.9\pm0.5$\rg\ of the central black hole.

\subsection{Self-consistent disk reflection}
\label{self_consis}

In all our previous fits a broad iron emission line has been shown to be robustly present above a thermal-disk and powerlaw-like continuum. In this section we use the reflection model \refbhb\ developed by \cite{refbhb} to self-consistently model both the thermal emission as well as the reflection features. Such analyses have previously been made for the stellar mass black hole in \gx\ and SWIFT~J1753.5-0127 where it was shown that the black holes are rotating with a dimensionless spin parameter of approximately $0.93\pm0.01$ and $0.76\pm0.15$ respectively (\citealt{reisgx,reis09spin}). A current shortcoming of the reflection model is that it assumes a single-temperature accretion disc. However, in \cite{reisgx} we showed that such an assumption does not have any significant effect on the inferred innermost radius of emission by: ({\rm i}) simulating a ``real'' disk with known inner radius and comparing it with the measured value using \refbhb\ and ({\rm ii}) comparing the results obtained with \refbhb\ with that of a thermal model, (KERRBB,  \citealt{kerrbb}), which includes relativistic smearing in a disc with radial temperature gradient. In both cases there were no indication of any strong systematic variation in the values obtained from the single temperature model possibly due to the the dominance of relativistic smearing over the intrinsic broadening due the nature of the multicolour disk.

\begin{figure}
{
 \rotatebox{270}{ 
 \resizebox{!}{8.cm} 
{\includegraphics[]{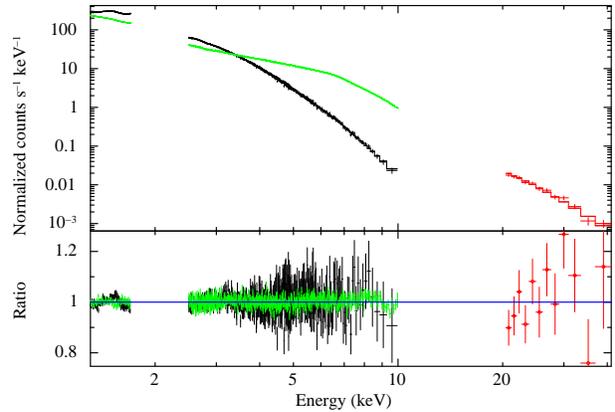}  
}}}
\caption{\suzaku\ and \xmm\ spectra of \j\ together with the ratio of the data to a model consisting of a self-consistent reflection component and a powerlaw (Model~4). The spectra have been binned for visual clarity only. \suzaku\ \xis\ and \pin\ are shown in black and red respectively. \xmm\ \epicpn\ spectrum is shown in green.         }
\label{fig_ratio_plot}

\end{figure}
\begin{figure*}
{
 \rotatebox{270}{ 
 \resizebox{!}{8cm} 
{\includegraphics[]{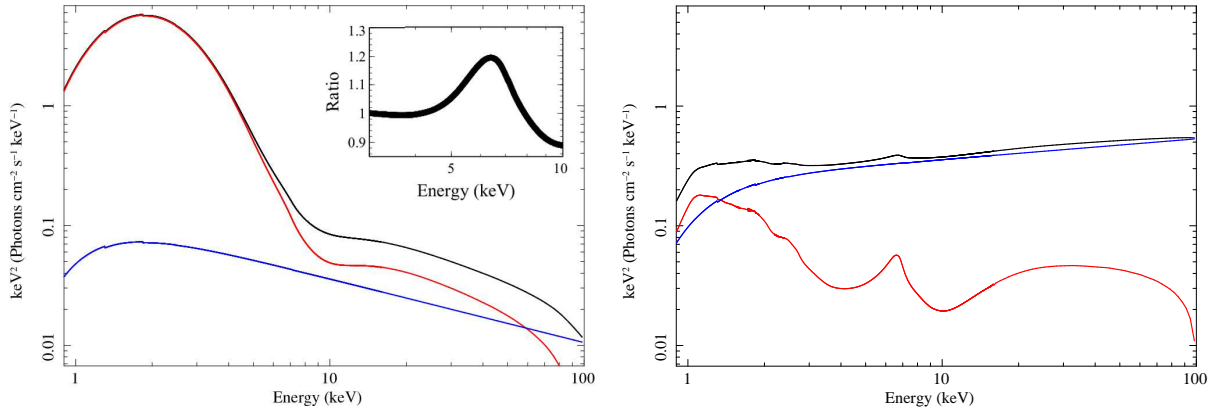} 
}}}
{
 \rotatebox{270}{ 
 \resizebox{!}{8cm} 
{\includegraphics[]{figure_eem_xmm.ps} 
}}}
\caption{Best-fit model for the \suzaku\ spectrum of \j\ in the intermediate state (Left) and \xmm\ spectrum in the low-hard state (Right). The total model is shown in black, with the powerlaw and \refbhb\ components shown in blue and red respectively. The inset in the left panel show the line profile obtained by dividing the current model with the best fit powerlaw plus \diskbb\ continuum. }
\label{fig_model}

\end{figure*}
The parameters of the \refbhb\ model are the number density of hydrogen in the illuminated surface layer, ${\it H}_{\rm den}$, the temperature of the blackbody heating the surface layers, the power-law photon index, and the ratio of the total flux illuminating the disc to the total blackbody flux emitted by the disc. To be able to directly compare the results with that obtained with the simple models (\S~\ref{simple}), we start by convolving the reflection component with the relativistic blurring kernel \kdblur, which is derived
from the same code by \citet{laor}. The power law index of \refbhb\ is
tied to that of the hard component and as before we constrain the inclination and inner disk radius to be the same between the observations. Note that we have now removed the upper-limit constraint on the inclination and allow the full parameter space to be explored. The model results in an excellent fit to the data with $\chi^{2}/\nu = 2579.3/2247$ (Model~4, see Fig.~\ref{fig_ratio_plot}). The parameters are shown in Table~\ref{table2}. Figure~\ref{fig_model} shows the model for both the \suzaku\ intermediate state (Left) and the \xmm\ low-hard state (Right) observations.

\begin{figure}
\centering
{
 \rotatebox{0}{ 
{\includegraphics[width=7.5cm, height=5.5cm]{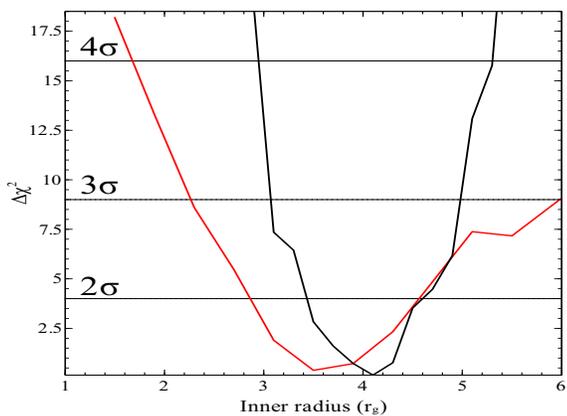}  
}}}
\caption{Goodness-of-fit versus inner radius for Model~1 (black) and Model~4 (red). The inner radii measured with these two distinct models are found to be consistent with each other. Taking the value obtained for the physically consistent model (Model~4; red) as a conservative indication of the inner extent of the accretion disk  we find \rin\ to be $3.7^{+0.6}_{-0.7}$\rg\ at the 90~per~cent confidence level ($\Delta\chi^2=2.71$ for one parameter of interest). The dotted lines indicate confidence intervals. }
\label{fig_steppar_plot}

\end{figure}

\begin{figure}
\centering
{
 \rotatebox{0}{ 
{\includegraphics[width=7.5cm, height=5.5cm]{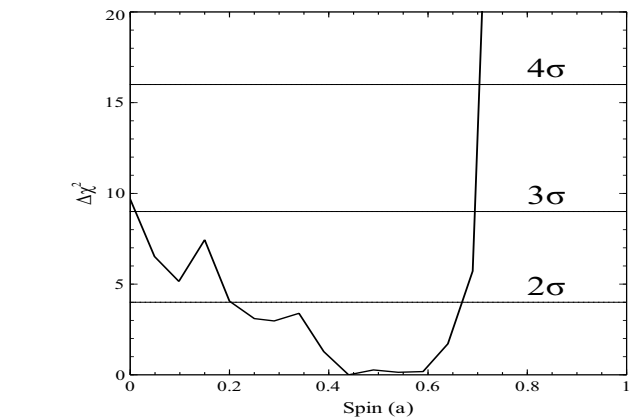}  
}}}
\caption{Goodness-of-fit versus spin for Model~5. It is clear that the black hole in \j\ is not maximally rotating, with a spin of $a=0.998$ being excluded at over 9$\sigma$. A non-rotating, Schwarzchild black is excluded at the $3\sigma$ level of confidence.   }
\label{fig_kerrconv}

\end{figure}

The results presented in Tables~1 and 2 suggest that the accretion disk in \j\ extends down to within $3.7^{+0.6}_{-0.7}$\rg\ of the central black hole. This constraint can be better appreciated in Fig.~\ref{fig_steppar_plot}, where the various confidences levels for \rin\ are shown as the dashed-lines in the $\chi^{2}$ plot. Also shown in Fig.~\ref{fig_steppar_plot} are the confidence levels for the value of \rin\ as obtained via the \laor\ line profile (Model~1). It is clear that the inner radius found by modelling the line profile is consistent with that obtained from the physically motivated and self-consistent reflection model thus confirming the robustness of results obtained from line-profile fitting in stellar mass black holes, AGNs and neutron stars (Stellar mass BH: \citealt{miller02j1650, miller04gx, miller09spin, blum09, reis09spin, hiemstra1652}; AGNs: \citealt{ Fabian02MCG, FabZog09, Fabianvaughan03, Zoghbi10,  miniutti09spin, schmoll09}; NS: \citealt{ bhattacharyya07, cackett08, cackett09, cackett10, disalvo09},  \citealt*{reisns}). Assuming that this radius is the same as the radius of marginal stability (\citealt{reynoldsfabian08}), our results implies that the black hole in \j\ is rotating with an intermediate spin parameter. To test this we replaced the convolution model \kdblur\ with the model \kerrconv\ (\citealt{kerrconv}). This convolution model is a fully relativistic code which enables the black hole spin to be fit as a free parameter and does not assume specific spin a priori. The parameters for this fit are listed in Table~\ref{table2} (Model~5). In this manner we find the spin of the black hole in \j\ to be $a=0.52^{+0.13}_{-0.16}$ at the 90~per~cent confidence level (Fig.~\ref{fig_kerrconv}) and place a strong upper limit of $a\lesssim0.7$ at greater than 5$\sigma$ .

\section{Summary and conclusions}
\label{discussion}

We have observed the black hole candidate \j\ on two separate occasions during the decay of its 2009 outburst (Fig.~\ref{fig_lightcurve}). By fitting the energy spectra we found that the earlier, \suzaku\ observation caught the source in the intermediate state with a luminosity of approximately 0.2$L_{Edd}$, whereas the subsequent \xmm\ observation found it to be in the canonical low-hard state with $L/L_{Edd}\approx 2\times10^{-2}$ assuming a 10\msun\ at a distance of 10\kpc. Both spectra shows the presence of a broad and asymmetric Fe emission line with equivalent width (EW) decreasing from 270\ev\ to 170\ev. The line profile is found to be remarkably similar between the two states (Fig.~\ref{fig_ratio_norm}) and robust to a variety of continua (Fig.~\ref{fig_ratio2pl}). Furthermore we find that a thermal-disk component is required in both states with the variation in flux following closely the $L\propto T^4$ relation. The requirement of both a thermal disk component as well as a strong (EW$>150$\ev) and asymmetric Fe emission line in the low-hard adhere to the strong observational criteria described in \cite{reislhs} and is further evidence that the accretion disk does not begin to recede at the onset of the low-hard state. Our result is in direct contrast with the truncated disk paradigm which suggest that the disk recedes at the outset of the low-hard state and is thereby replaced by advection or magnetically dominated accretion flows.

By modelling the full reflection features present in the spectra of \j\ with a self-consistent reflection model, we have shown in Fig.~\ref{fig_steppar_plot} that the inner disk extends down to $3.7^{+0.6}_{-0.7}$\rg\ at the 90~per~cent level of confidence. Such a joint fit to the multiple states with a common model allows for the explicit treatment of variations in the ionization and/or structure of the accretion disk whilst at the same time reducing the errors on intrinsic physical parameters such as the inner-disk inclination and black hole spin (\citealt{miller08gx, reisgx, miller09spin}). We note also that similar values for \rin\ are found when fitting the Fe-\ka\ line profile with a variety of phenomenological models, thus confirming the robustness of inner radii measurements via such reflection features. By using a fully-relativistic convolution code (\kerrconv) acting on a self-consistent reflection model (\refbhb), we formally constrained the spin of the black hole in \j\ to be $a=0.52^{+0.13}_{-0.16}$ at the 90~per~cent level of confidence, thus ruling out a non-rotating, Schwarzchild black hole at greater than the 3$\sigma$ confidence level and a maximally rotating black hole at more than the $9\sigma$ level of confidence (Fig.~\ref{fig_kerrconv}).
   
The clear spectral differences between the two states  (see Fig.~\ref{fig_model}) can potentially be ascribed to physical changes in the disk properties. In the intermediate state the disk contributes mostly towards the heating of the atmosphere whereas in the low-hard state this is achieved mostly by flux from the comptonzing region (see Table~\ref{table2}). This difference is partially due to the much lower thermal flux in the low-hard state, but it may also reflect an increase in the accretion power being funnelled into the corona or jet. The increase in hydrogen number density in the low-hard state as compared to the intermediate state, together with the lack of radial variation in the disk is likely attributed to changes in the atmosphere of the inner disk indicating that the disk is more vertically-extended in the intermediate state. Changes in the physical properties of the inner disk has previously been proposed as viable alternative to disk truncation (\citealt{merlonifabianross00}) and has recently been invoked to partly explain the variable behaviour of \gx\ in the low-hard state (\citealt{wilkinsonuttley09}). The  decrease in the equivalent width of the iron \ka\ line between the states is characteristic of a decrease in the reflection fraction ($R$, \citealt{George91}). This behaviour is usually seen in stellar mass black holes and can be explained without invoking disk truncation by assuming that either the inner disk becomes fully ionized (\citealt{rossfabianyoung99}) or/and that the corona  is moving relativistically away from the disk  (\citealt{beloborodov99}).

The natal spin of a stellar-mass black holes produced via single collapse events is often thought to be be $\lesssim0.75-0.9$ depending on the type of the supernova or GRB that preceded its formation and the physical structure (magnetic field, angular momentum, metallicity, etc) of the progenitor star (see e.g. \citealt{shibata02, hegerwoosley02,gammie04}). However, subsequent accretion process has been shown to spin-up the central black hole to values $>0.9$ (\citealt{gammie04, volonteri05}) under certain conditions. The intermediate value for the spin parameter found here suggest that the black hole in \j\ has not changed its natal spin by a significant amount.

\section{Acknowledgements}
RCR would like to thank the anonymous referee and the Science and Technology Research Council (STFC). ACF thanks the Royal Society. EMC and CSR gratefully acknowledges support provided by NASA through the Chandra Fellowship Program and the US National Science Foundation under grant AST 06-07428 respectively.

\bibliographystyle{mnras}

\bibliography{ref}
\end{document}